\documentclass[nouppercase]{kaia}
\usepackage{cite}
\usepackage{amsmath}
\usepackage{tabularx}
\usepackage{algorithm}
\usepackage{algpseudocode}
\usepackage{mathtools}
\usepackage{graphicx} % Required for including images

\title{PAS: Partial Additive Speech Data Augmentation Method for Noise Robust Speaker Verification}
\author{Wonbin Kim}{a}
\author{Hyun-seo Shin}{a}
\author{Ju-ho Kim}{a}
\author{Jungwoo Heo}{a}
\author{Chan-yeong Lim}{a}
\author{Ha-Jin Yu}{a}
\affiliation{School of Computer Science, University of Seoul, Republic of Korea, rst0070@uos.ac.kr, gustjtls123@naver.com, wngh1187@naver.com, jungwoo4021@gmail.com,  cksdud585@naver.com, hjyu@uos.ac.kr}{a}

\begin{document}
\maketitle
\let\thefootnote\relax\footnotetext{This work was supported by the National Research Foundation of Korea(NRF) grant funded by the Korea government. (MSIT) (2023R1A2C1005744)}
\begin{abstract}
Background noise reduces speech intelligibility and quality, making speaker verification (SV) in noisy environments a challenging task.
To improve the noise robustness of SV systems, additive noise data augmentation method has been commonly used.
In this paper, we propose a new additive noise method, partial additive speech (PAS), which aims to train SV systems to be less affected by noisy environments.
The experimental results demonstrate that PAS outperforms traditional additive noise in terms of equal error rates (EER), with relative improvements of 4.64\% and 5.01\% observed in SE-ResNet34 and ECAPA-TDNN.
We also show the effectiveness of proposed method by analyzing attention modules and visualizing speaker embeddings. 
\end{abstract}

\section{Introduction}
Speaker verification (SV) is a task verifying whether the speaker of a given speech segment matches the enrolled speaker.
In recent years, deep learning-based SV systems\cite{snyder2017deep, snyder2018x} have shown significant improvements compared to traditional methods\cite{reynolds2000speaker, dehak2010front}, achieving state-of-the-art performance\cite{desplanques2020ecapa, chung2020in}.
The process of deep learning-based SV typically involves extracting speaker feature vectors (speaker embeddings) from the input speeches and comparing these vectors to calculate their similarity.

%However
Despite the great advances in deep learning-based SV systems, performance degradation in noisy environments remains one of the challenges in SV task, because the background noise can mask or distort important spectral and temporal characteristics of the speech for speaker verification\cite{wolfel2009distant}. 
To address this challenge, researchers have developed various techniques\cite{shon2019voiceid, park2019specaugment, kim2022extended}. 

In particular, additive noise method\cite{medina2003robust, snyder2018x} is one of the popular methods to make SV systems robust in noisy environments.
The method trains the SV system to extract speaker characteristics in scenarios where noise affects the information. 
Most studies\cite{cai2020fly, desplanques2020ecapa, yu2021cam} have demonstrated effectiveness of the additive noise method described in \cite{snyder2018x} which involves adding noise audio to speech audio throughout the entire duration.

Although the training method to extract speaker information from noisy environments has been successful, 
it is also important to minimize the influence of noise-dependent features to the information. 
This is because capturing speaker characteristics is the goal of the SV task\cite{snyder2017deep}, rather than focusing on environmental factors.
To achieve this, it is important to enhance the SV system's ability to distinguish between speech and background noise. 
Attention mechanism is one of the popular approach to achieve that goal. 
Attentive Statistics Pooling (ASP) uses attention mechanism to emphasize frame-level features that are important for extracting speaker embeddings.
Squeeze-and-Excitation (SE) Block uses attention-based method to help convolutional neural network (CNN) focus on important features.

Inspired by the attention mechanisms that encourage SV systems to extract speaker-relative information, we propose a novel data augmentation method, partial additive speech (PAS).
PAS aims to promote SV systems to extract speaker information which is noise-independent. 
Unlike the traditional additive noise method, PAS incorporates noise-only segments into the training data.
We expect that training SV systems with this type of data will contribute to achieving the goal by emphasizing irrelevance between speaker features and noise-dependent features.
Moreover, incorporating PAS can further enhance the attention mechanisms of SV systems.

In this paper, we show that PAS is an effective data augmentation technique in comparison to the traditional additive noise method.  
Through experiments, we examine the impact of PAS on the attention mechanism by removing attention modules from SV systems. 
By comparing the speaker embedding distribution of systems trained using PAS and the traditional method, we have observed that PAS enhances the ability of SV systems to distinguish between noise and speech.

\begin{figure}[t]
  \begin{center}
    \centering
    \vspace{-0.2cm}
    \includegraphics[width=\linewidth]{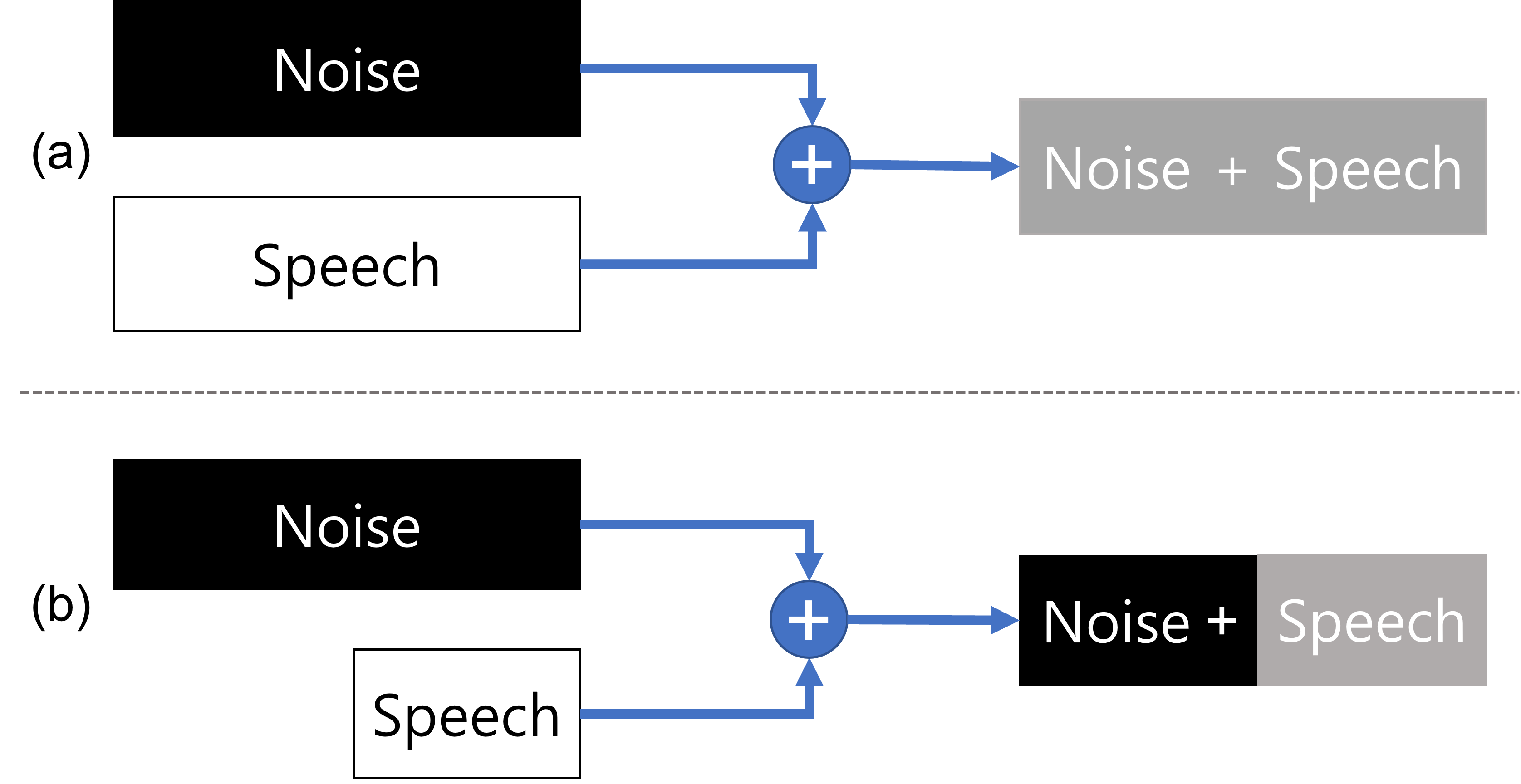} 
    \caption{
        Diagrams of two different data augmentation methods implementing additive noise. 
        (a): traditional additive noise method, (b): proposed PAS method. 
        + symbol denotes element-wise sum.
    }
    \label{fig:diagrams}
  \end{center}
  \vspace{-0.4cm}
\end{figure}

\section{Related Work}
In this section, we overview several related research on traditional data augmentation methods for noise-robust SV systems and attention mechanisms.
In the section\ref{section:Result}, we analyze the impact of the PAS method on attention mechanisms for SV systems.

\textbf{Additive noise} is a representative data augmentation method in SV task to simulate real-world scenarios where the speech features are degraded by background noise\cite{snyder2018x}. 
A conventional approach is to incorporate noise audio of the same duration as the speech audio, as shown in Figure\ref{fig:diagrams}(a).
It aims to mimic the effect of noise in real-world conditions and has been shown to enhance the robustness of SV systems in noisy environments\cite{medina2003robust, snyder2018x}.

\textbf{Attentive Statistics Pooling (ASP)}\cite{okabe2018attentive}  is a pooling method that employs an attention mechanism to generate a weighted mean and standard deviation for the utterance-level features extracted from the frame-level features.
The basic idea behind ASP is that the attention mechanism can function to calculate the importance of each frame to speaker verification, e.g., to give noisy frames small weights. 
ASP improved SV performance by empowering the SV system to focus on the crucial frames to extract speaker embeddings.

\textbf{Squeeze-and-Excitation (SE) Block}\cite{hu2018squeeze} consists of attention mechanism-based layers that aim to enhance the expressiveness of convolutional neural networks (CNN).
The objective of the SE-Block is to extract crucial information by modeling the interdependencies among the channels of feature maps.
This architecture involves two key operations: squeeze and excitation.

During the squeeze operation, global average pooling is applied to each channel of the feature map capturing the importance of each channel. 
By using the statistics extracted from squeeze operation, the excitation operation then emphasize informative features and suppress less useful ones by attention mechanism. 

\begin{table*}[t!]
\resizebox{\textwidth}{!}{
    \centering
    \begin{tabular}{c | c | c c c | c c c | c c c | c c c}
        \hline
        Noise type              & SNR       &  \multicolumn{12}{c}{Equal Error Rate(EER) \%} \\
        
        \hline
        \multicolumn{2}{c|}{Framework} & \multicolumn{3}{c|}{SE-ResNet34} & \multicolumn{3}{c|}{SE-ResNet34 w/o Attn} & \multicolumn{3}{c|}{ECAPA-TDNN} & \multicolumn{3}{c}{ECAPA-TDNN w/o Attn} \\
        
        \hline
        \multicolumn{2}{c|}{Data augmentation}   &  ×  & TAN & PAS & ×   & TAN & PAS & ×   & TAN & PAS & ×   & TAN & PAS \\
        
        \hline
        \multicolumn{2}{c|}{
            Original test set $T$}          & 3.96  & 3.65      & \textbf{3.35}      &   4.57     & 3.92      & \textbf{3.84}      &  4.18         & 3.46      &   \textbf{3.32}    & 4.91      &   3.75    &   \textbf{3.71} \\
        
        \hline
        \multirow{5}{*}{Babble} &   0              & 27.68 &    12.14   &   \textbf{11.99}   & 29.67 &  13.15   &   \textbf{12.68}   & 26.18 &   13.01   &   \textbf{11.86}   & 27.63 &  13.72   &   \textbf{13.26} \\
                                &   5              & 14.95 &   6.56 	&   \textbf{6.24} 	& 17.25 &  7.18 	&   \textbf{7.07} 	& 13.94 &   7.09 	&   \textbf{6.83} 	& 15.51 &  \textbf{7.54} 	&   7.66 \\
                                &   10             & 8.36 &   5.08 	&   \textbf{4.77} 	& 9.98 &  5.35 	&   \textbf{5.22} 	& 8.15 &  5.03 	&  \textbf{4.91} 	& 9.26 &  \textbf{5.39} 	&   5.40 \\
                                &   15             & 5.98 &   4.40    &   \textbf{4.04}    & 6.83 &  4.52    &   \textbf{4.47}    & 5.81 &  4.21    &   \textbf{4.03}    & 6.40 &  \textbf{4.40}    &   4.42 \\
                                &   20             & 4.81 &  4.01    &   \textbf{3.66}    & 5.28 &   \textbf{4.15}    &   4.16    & 4.83 &  \textbf{3.63}    &   3.65    & 5.55 &  \textbf{4.00}    &   \textbf{4.00} \\
        \hline
        \multirow{5}{*}{Music}  &   0              & 20.33 &   7.97    &   \textbf{7.82}    & 23.63 &  8.58    &   \textbf{8.28}    & 20.88 &  9.92    &   \textbf{9.02}    & 22.61 &  10.53   &   \textbf{10.17} \\ 
                                &   5              & 11.68 &   5.62 	&   \textbf{5.60} 	& 13.67 &  6.24 	&   \textbf{5.99} 	& 12.94 &  6.44 	&   \textbf{6.11} 	& 13.90 &  6.95 	&   \textbf{6.86}  \\
                                &   10             & 7.86 &   4.60 	&   \textbf{4.50} 	& 9.08 &  4.92 	&   \textbf{4.85} 	& 8.30 &  4.85 	&   \textbf{4.82} 	& 8.97 &  5.25 	&   \textbf{5.22}  \\
                                &   15             & 5.66 &   4.24 	&   \textbf{3.82} 	& 6.22 &  4.45 	&   \textbf{4.40} 	& 5.74 &  3.99 	&   \textbf{3.87} 	& 6.41 &   \textbf{4.33} 	&   4.40  \\ 
                                &   20             & 4.79 &   3.97 	&   \textbf{3.57} 	& 5.32 &  4.14 	&   \textbf{4.10} 	& 4.84 &  3.81 	&   \textbf{3.77} 	& 5.55 &  4.17 	&   \textbf{4.05}  \\
        \hline
        \multirow{5}{*}{Noise}  &   0              & 19.98 &   8.61 	&   \textbf{8.43}	& 22.29 &  9.22 	&   \textbf{8.82} 	& 22.39 &  9.75 	&   \textbf{9.06} 	& 24.66 &  10.42 	&   \textbf{10.00} \\
                                &   5              & 12.74  &  5.97 	&   \textbf{5.93} 	& 14.25  & 6.78 	&   \textbf{6.59} 	& 13.92 &  6.98 	&   \textbf{6.52} 	& 15.29 &  7.24 	&   \textbf{6.98}  \\
                                &   10             & 8.53 &   5.10 	&   \textbf{4.82} 	& 9.25 &  5.59 	&   \textbf{5.32} 	& 9.15 &  5.15 	&   \textbf{5.13} 	& 10.23 &  5.67 	&   \textbf{5.61}  \\
                                &   15             & 6.51 &   4.54 	&   \textbf{3.99} 	& 6.93 &  4.82 	&   \textbf{4.64} 	& 6.57 &  4.43 	&   \textbf{4.21} 	& 7.16 &  4.64 	&   \textbf{4.64}  \\ 
                                &   20             & 5.02 &   4.05 	&   \textbf{3.75} 	& 5.69 &  \textbf{4.32} 	&   4.34 	& 5.37 &  4.03 	&   \textbf{3.88} 	& 6.13 &  4.28 	&   \textbf{4.05}  \\ 
        \hline
        \multicolumn{2}{c|}{Average EER}           & 10.55 &  5.66    &   \textbf{5.39} 	& 11.87 &  6.08 	&   \textbf{5.92} 	& 10.82 &  5.99 	&   \textbf{5.69} 	& 11.89 &  6.39 	&   \textbf{6.28}  \\
        \hline
    \end{tabular}
}
\\ \\
\caption{
    Experimental results evaluated on the VoxCeleb1-Original trial and noise scenarios synthesized with the MUSAN corpus at different SNRs. 
    Attn and TAN refer to the attention mechanism and traditional additive noise method, respectively.
}
\label{table:result}
\end{table*}

\section{Proposed Method}
We propose a novel data augmentation method, PAS, for enhancing the effectiveness of traditional additive noise method. 
Inspired by attention mechanisms, the goal of PAS is to improve the SV system's ability to extract speaker information from noisy environments while minimizing the environmental factors.
PAS generates training data which has two different parts, background noise only and speech degraded by the background noise. 
We expect that this kind of data will enhance the ability of SV systems for extracting noise-independent speaker information. 

The augmentation process is conducted by Algorithm\ref{alg:cap}. 
\begin{algorithm}
\caption{Applying PAS to a mini-batch}
\label{alg:cap}
\textbf{Input:} mini-batch $X$, noise source $N$, hyper parameters $L_n, L_{s\_min}, SNR_{min}$ and $SNR_{max}$.\\
$L_n$ is the fixed length for noise audio. 
$L_{s\_min}$, the minimum length required for each speech audio. 
$SNR_{min}$, the minimum value of the Signal-to-Noise Ratio (SNR). 
$SNR_{max}$, the maximum value of the SNR.  \\
\begin{algorithmic}
\State $S \gets [\,]$  \Comment{Empty list}
\For{$x$ in $X$}
    \State $n \gets$ randomly choose a noise audio from $N$
    \State $n \gets sample\_segment(n, L_n)$  \Comment{Sample a random}
    \State \hfill segment with duration $L_n$
    \State $L_s \sim U(L_{s\_min}, L_n)$  \Comment{Randomly choose}
    \State \hfill an increment value 
    \State \hfill from uniform distribution
    \State $x \gets sample\_segment(x, L_s)$
    \State $SNR \sim U(SNR_{min}, SNR_{max})$
    
    \State $n \gets adjust\_amplitude(n, SNR)$  \Comment{adjust}
    \State \hfill amplitude of $N$
    \State \hfill to match SNR
    
    \State $P_s \sim U(0, L_n - L_s)$  \Comment{Position for}
    \State \hfill synthesizing speech
    \\
    \State $seg_1 \gets clip(n, 0, P_s - 1)$  \Comment{Clip noise}
    \State \hfill within $[0, P_s - 1]$
    
    \State $seg_2 \gets x + clip(n, P_s, P_s + L_s  - 1)$  \Comment{Add}
    \State \hfill  noise and speech
    
    \State $seg_3 \gets clip(n, P_s + L_s, L_n - 1)$

    \State $segments \gets concatenate(seg_1, seg_2, seg_3)$

    \State $S \gets append(S, segments)$
\EndFor
\State \Return $S$

\end{algorithmic}
\end{algorithm}

After synthesizing, each augmented audio has the following parts: 
\begin{itemize}
    \item background noise only: $[0, P_s - 1]$, $[P_s + L_s, L_n - 1]$
    \item speech with the background noise: $[P_s, P_s + L_n - 1]$
\end{itemize}

\section{Experiment Settings}
\subsection{Datasets}
In our experiments, we used the VoxCeleb1 dataset\cite{nagrani2017voxceleb}. 
The VoxCeleb1 development set consists of 148,642 utterances from 1,211 speakers and the test set contains 4,874 utterances from 40 speakers. 

To simulate noisy environments, we used the MUSAN corpus\cite{snyder2015musan}. 
We divided it into two non-overlapping parts, one for training and one for testing. 

\subsection{Data augmentation}
To demonstrate the effectiveness of the proposed method, we conducted data augmentation using two different approaches: PAS and traditional additive noise. 

While training, each method were applied to three out of every four utterances. 
For PAS training, the hyper parameters of PAS were $L_n\,=3.2s\, ,L_{s\_min}\,=\,1s ,SNR_{min}\,=\,0$ and $SNR_{max}\,=\,20$.
Traditional additive noise was applied to each training utterance with randomly selected SNR value from a uniform distribution of $[0, 20]$. 

For the evaluation under noise scenarios, we applied traditional additive noise to VoxCeleb1 test set. We synthesized the test set for each noise category with SNRs in the set \{0, 5, 10, 15, 20\}.

\subsection{Backbone frameworks}
In our experiments, we used two CNN-based SV backbone frameworks to validate our proposed method: SE-ResNet34\cite{hu2018squeeze, chung2020in} and ECAPA-TDNN\cite{desplanques2020ecapa}. 
These frameworks are based on 2-dimensional and 1-dimensional convolution layers, respectively.

We also built non-attentional versions of each framework to investigate the interaction between the attention mechanism and PAS by removing the attention modules.
To accomplish this, we removed SE-Module from the framework and replaced ASP with statistics pooling.

\subsection{Implementation Details}
We conducted performance evaluation on the VoxCeleb1-Original trial which is one of the three trial pairs of VoxCeleb1 dataset. 
We used 80-dimensional log Mel-spectrograms for input features. 
These spectrograms were obtained using a 1024-point FFT with a hamming window of width 25ms and a hop size of 10ms.
To construct mini-batches, we randomly cropped 3.2 seconds of audio from each utterance, resulting in a total of 100 utterances. 
For each utterance, there is a 0.75 probability of mixing it with noise audio, and the selection of the noise audio is randomized independently for each utterance.
The adam optimizer\cite{kingma2014adam} was utilized with a learning rate of 0.001 and a weight decay of 0.0001. 
The learning rate was reduced by 6\% for every epoch, and each framework was trained for 100 epochs.
We employed AAM-Softmax\cite{deng2019arcface, xiang2019margin} loss function with different margins and scales for each framework. 
For training the SE-ResNet34, we set the margin to 0.1 and the scale to 15, whereas for the ECAPA-TDNN, we set 0.3 and 15 respectively. 

\begin{figure}[t]
  \begin{center}
    \centering
    \vspace{-0.2cm}
    \begin{minipage}{\linewidth}
    \includegraphics[width=\linewidth]{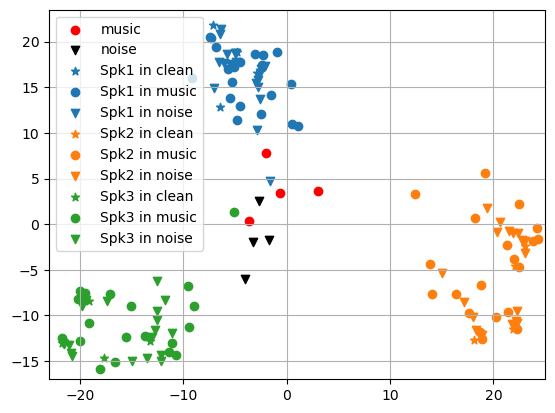}
    %\captionof*{.}{(a) traditional additive noise}
    \centering (a) traditional additive noise
    \end{minipage}
    \begin{minipage}{\linewidth}
    \includegraphics[width=\linewidth]{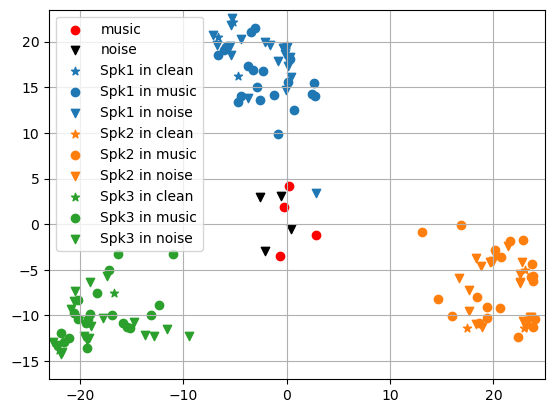} 
    \centering (b) PAS
    \end{minipage}
    
    \caption{
        Distribution of speaker embeddings after PCA, extracted from ECAPA-TDNN.
        (a) is from SV system trained with traditional additive noise method, and (b) is from the system trained with PAS.
        Spk1, spk2 and spk3 denote three different speakers.
        Noise scenarios are music and noise. 
        Clean denotes scenario without additional noise.
    }
    \label{fig:embedding1}
  \end{center}
  \vspace{-0.4cm}
\end{figure}

\section{Results and Analysis}
\label{section:Result}
In this section, we analyze the traditional additive noise and our proposed method.  
Then, based on the experimental results, we investigate the impact of PAS on the attention mechanism. 
To gain deeper insights into the impact of PAS on the SV system's ability to distinguish between different speakers in noisy environments, we visualize the speaker embeddings. 
These embeddings are extracted from both a PAS-trained SV system and a traditional additive noise-trained SV system. 

\subsection{Analysis of PAS for attention and non-attention SV systems}
Table \ref{table:result} presents the results obtained from three data augmentation configurations, helping a comparison of the effectiveness between the traditional additive noise method and PAS.
We also compare the performance of each framework for both attention-version and non-attention-version to discover the impact of PAS on the attention mechanism.
  
For systems without attention, the PAS method has a relatively improved equivalent error rate (EER) compared to the traditional additive noise method.
Specifically, the relative improvements in EER were observed as 2.63\% and 1.82\% for non-attention SE-ResNet34 and non-attention ECAPA-TDNN.
It is notable that PAS performs better, even though the traditional additive noise and test set are generated in the same way. 

For systems with attention, the PAS method exhibited relative improvements in EER of 4.64\%, 5.01\% over the traditional additive noise method. 
This notices that frameworks incorporating an attention mechanism showed greater improvement rates. 
This suggests that PAS enhances the attention mechanism to achieve it's goal discerning the relative importance of input features in speaker verification. 

Moreover, the fact that the performance was enhanced on the systems without attention mechanism also shows that PAS helps to improve the SV system's ability to distinguish speech from background noise. 

\subsection{Distribution of speaker embeddings in noisy scenarios}
To analyze the impact of training with the PAS method on the behavior of a SV system in noisy environments, we visualized speaker embeddings using principal component analysis (PCA) \cite{abdi2010principal}. 
Additionally, we extracted embeddings solely from the noise audio used in constructing the noisy environments to watch the ability of SV systems extracting noise-independent information. 
For this experiment, we selected four utterances from each of three different speakers in the VoxCeleb1 test set. 

To create the noise environment, we incorporated audio from the noise and music categories of the MUSAN dataset, varying the signal-to-noise ratio (SNR). 
We utilized the pretrained ECAPA-TDNN framework the outcomes of which are outlined in Table \ref{table:result}. 

In Figure \ref{fig:embedding1}, (a) illustrates the speaker embeddings trained with traditional additive noise method, while (b) represents the embeddings trained with the PAS method. 
In both cases, we observe distinct separation between the embeddings of the three speakers and the embedding of the noise. 
However, in (a), the distribution range of embeddings for the same speaker appears relatively wider, and there is overlap between the noise embeddings and the speaker embeddings.
We can also observe that the distribution of noise embeddings is wider than PAS trained.

These results imply that training with the PAS can improves the system's ability to extract speaker information which is more noise-independent. 
Consequently, PAS contributes to enhance speaker verification performance. 

\section{Conclusion}
Additive noise method has been commonly used to enhance the robustness of Speaker Verification (SV) systems in noisy environments and has shown its effectiveness. 
In this paper, we aimed to improve traditional additive noise method by promoting the extraction of speaker information independent of background noise.
To achieve this goal, we proposed the Partial Additive Speech (PAS) method, which partially synthesizes noise audio with speech audio of shorter or equal length to the noise audio.
Our experiments on the VoxCeleb1 dataset and MUSAN dataset demonstrated that the PAS method outperforms the traditional additive noise method in equal error rate(EER). 
Furthermore, our experiments investigating the interaction between PAS and the attention mechanism revealed that PAS enhances the SV system's discriminative capabilities for input features. 
The visualization of speaker embeddings also indirectly indicates that PAS  increases the system's ability to extract noise-independent speaker information.

\bibliographystyle{plain}
\bibliography{references}

\thispagestyle{empty}
\end{document}